\documentclass{article}[12pt]
\usepackage{amsmath}
\usepackage{amsfonts}
\usepackage{amssymb}
\usepackage{amsthm}
\usepackage{graphicx}
\usepackage{natbib}
\usepackage{bbm}
\usepackage{contour}
\usepackage{comment}
\usepackage[margin=1.25in]{geometry}
\contourlength{1pt}
\contournumber{40}
\usepackage{pgf,tikz}
\usepackage[hidelinks]{hyperref}

\theoremstyle{remark}

\defcitealias{NAP25303}{National Academies 2019}
\begin{document}
	\title{From one environment to many: The problem of replicability of statistical inferences} 
	
	\author{James J.~Higgins, Michael J.~Higgins, Jinguang Lin\footnote{Department of Statistics, Kansas State University}}
	\date{\small{\vspace{.4in} \today }}
	
	\maketitle\textbf{}

\begin{abstract}
Among plausible causes for replicability failure, one that has not received sufficient attention is the environment in which the research is conducted. 
Consisting of the population, equipment, personnel, and various conditions such as location, time, and weather, the research environment can affect treatments and outcomes, and changes in the research environment that occur when an experiment is redone can affect replicability. 
We examine the extent to which such changes contribute to replicability failure. 
Our framework is that of an initial experiment that generates the data and a follow-up experiment that is done the same way except for a change in the research environment. 
We assume that the initial experiment satisfies the assumptions of the two-sample $t$-statistic and that the follow-up experiment is described by a mixed model which includes environmental parameters. 
We derive expressions for the effect that the research environment has on power, sample size selection, $p$-values, and confidence levels. 
We measure the size of the environmental effect with the environmental effect ratio (EER) which is the ratio of the standard deviations of environment by treatment interaction and error. 
By varying EER, it is possible to determine conditions that favor replicability and those that do not. 
\end{abstract}

\noindent%
\textbf{Key Words}: replicability, environment by treatment interaction, treatment effect size, broad-inference $p$-value

\section{Introduction}
\label{sec:Intro}

The replicability crisis---the phenomenon that conclusions from many scientific studies are unable to be verified in follow-up studies---was dramatically brought to the forefront when researchers at the biotech company Amgen attempted to replicate 53 well-regarded pre-clinical studies only to find that just 6 supported the original conclusions~\citep{begley2012raise}. Failures to replicate have been shown to occur across all scientific domains~\citep{begley2015reproducibility}. A survey of Nature readers found that about 70$\%$ of scientists have failed to replicate other researchers' experimental results, and more than 50$\%$ have failed to replicate results of their own studies~\citep{baker2016there}. 

The \citet{NAP25303} define \textit{replicability} as ``obtaining consistent results across studies aimed at answering the same scientific question, each of which has obtained its own data,'' and recently addressed the replicability crisis from the point of view of science in general. 
The statistics community came out with a special issue of The American Statistician to deal with statistical issues~\citep{ronspaper}. 
Much of the literature has pointed to familiar culprits behind the replicability crisis: $p$-hacking, insufficient power, inappropriate analyses given the data, variability of $p$-values, overstated evidence, and so on~\citep{ioannidis2009repeatability,boos2011p,allison2018reproducibility, bello2018invited, gibson2020role}.
Concerns about replicability have led to an overall distrust of $p$-values and claims of statistical significance. 
It has progressed to the point where practitioners have recommended removing such terms and phrases completely~\citep{woolston2015psychology,ronspaper}.

While the aforementioned violations of recommended statistical practice are common reasons why an experiment cannot be replicated, we shed light on another concern, namely the \textit{research environment} and its effect on the treatments and outcomes. The research environment encompasses the totality of conditions under which an experiment is done. This includes the population, if any, the protocol, personnel, and the prevailing conditions such as the weather, location, equipment, and time of day or year. It also includes unrecognized random factors that could systematically bias outcomes in unknown ways. 

The research environment is unique to each experiment; it changes when the experiment is repeated. 
The magnitude of the change may be small, as in a tightly controlled laboratory experiment or large, as in an agricultural field trial. 
Furthermore, if there is environment by treatment interaction, which appears to be common in research~\citep{berliner2002comment, Kafkafi:2005, kafkafi2017addressing, kafkafi2018reproducibility, snape2007dissecting}, treatments with large effects in one environment could have small effects in another; moreover, such discrepancies are not the result of misuse of statistical methods but occur because of inherent difficulties in doing an experiment exactly the same way twice. 


The effect of a changing research environment on replicability may be inferred by performing the same experiment in several randomly selected environments and applying a mixed model analysis to assess treatment effects. 
Such mixed-model analysis has been a staple of fields including agriculture, biology, and psychology~\citep{littell1996sas, milliken2009analysis, Kafkafi:2005, cronbach1963theory, cronbach1972dependability, shavelson1989generalizability}.
However, this approach is limited because it is often impractical to do a study more than once due to costs, time, or lack of incentives for carrying out replication studies~(\citetalias[p.~137--138]{NAP25303},~\citealt{koole2012rewarding, lundwall2019changing}).  
Meta-analysis may be appropriate when the same treatments are studied in several environments, usually with non-identical experiments, and the desire is to come up with estimates of treatment effects that are better than one could get with a single experiment~\citep{lipsey2001practical, borenstein2011introduction}. 


Typically, measures of replicability do not account for a changing environment.  For example, \citet{goodman1992comment} proposed the ``probability of repeating a significant result in the same direction.'' It is computed after an initial experiment is done and assumes that the follow-up experiment, if it were to be done, would be done under the identical conditions of the initial experiment. \citet{trafimow2018priori} computed a “probability of replication” based on a criterion of closeness of sample means to population means in the initial and follow-up experiments. It can be computed before any experimentation is done, but it also does not account for random environmental effects except for sampling error.  



In practice, replicability is often just an assumption based on the scientific tenet that experiments done the same way will produce the same results. However, the replicability crisis has brought this assumption into question.  Thus, it would be desirable to have probability-based inferences that can assess whether an assumption of replicabilty is reasonable in light of potential changes in the research environment that may occur in future experiments.

\citet{kafkafi2017addressing} dealt with this problem in the context of single-laboratory genetic experiments. 
They assumed that observations from a single laboratory are governed by the same mixed model that would apply to multiple-laboratory experiments. 
They estimated genotype-by-laboratory interaction from previous studies and combined this information with the data from the present experiment to adjust the variance of the difference of means. 
They used the $t$-statistic with Satterthwaite’s approximation for degrees of freedom to obtain $p$-values and confidence intervals that are "adjusted" for changing environments.

We propose a conceptual framework for thinking about replicability that makes it possible to pose and answer a broader array of questions than could be addressed with just a single number such as ``probability of replicability.'' 
Replicability in our framework is about the consistency of the properties of inferential procedures as they are applied to initial and follow-up experiments. 
The initial experiment is where the data are taken; the follow-up experiment is a hypothetical construct that enables us to ask and answer ``what if'' questions about replicability in a changing research environment. 
For instance, if the initial experiment has a desirable power for a given pre-specified effect size, how would the power be affected by the change in environment in a follow-up experiment?  
Couching replicability in terms of properties of inferential procedures rather than experimental outcomes is implicit in the popular recommendation to use large $n$ and small $p$ as a means to deal with replicability problems~\citep{singh2017big}.  
Here, the properties that are assumed to be replicable are small probabilities of Type I and Type II errors. 
Ironically, we are able to use our conceptualization to show serious flaws in this recommendation (see Section~\ref{subsec:ex2}).

We measure the size of the environmental effect with a dimensionless parameter that we call the environmental effect ratio EER, which is the ratio of the standard deviations of environment-by-treatment interaction and the experimental error. 
We obtain exact distributions of the statistical quantities of interest by assuming that EER is known. 
We further assume that information is available that would enable the researcher to place reasonable bounds on EER and therefore to be able to place reasonable bounds on quantities that depend on it. 

Additionally, we show that by varying EER, it is possible to examine the sensitivity of methods to a changing research environment. 
We apply this approach to power, sample size selection, $p$-values, and confidence levels. 
In so doing we are able to determine conditions that are favorable for replicability and those that are not. 
In addition to EER, these conditions depend on sample size, effect size, and in the case of hypothesis testing, nominal significance levels.

Section 2 has the mathematical model and exact distributional properties of the t-statistic under the mixed model. Section 3 considers the effect of the research environment on power, relative efficiency, and sample size selection. In Section 4 we derive broad-inference p-values and confidence intervals, which are functions of EER, and show how these can be used to help researchers make inferences across multiple environments from data taken from just one environment. Section 5 gives plausible values of EER from data that have been taken in multi-environment experiments. Section 6 has summary and conclusions.

\section{Model}
\label{sec:Model}

We consider an experiment with two treatment conditions $r \in (1,2)$.
Observations from the initial experiment are assumed to follow the model
\begin{equation}
\label{eq:initialmodel}
Y_{rj} = \mu_r + \epsilon_{rj},\;\;\ r = 1,2, \;j = 1,\ldots, n_r
\end{equation}
where $\mu_r$ is the mean of the $r$th treatment in the environment in which the initial experiment is run, $n_r$ is the number of units assigned to treatment $r$, and the $\epsilon_{rj}$'s are independent and identically distributed (iid) $N(0, \sigma^2_e)$ random variables.  We are interested in inferences for $\mu_1 - \mu_2$.

In a follow-up experiment, the changing research environment is assumed to affect the responses all in the same way or in ways unique to each treatment. We express this with the mixed model
\begin{equation}
\label{eq:followupmodel}
Y_{rj} = \mu_r + \theta + \delta_r + \epsilon_{rj},\;\;\ r = 1,2, \;j = 1,\ldots, n_r.
\end{equation}
The $\mu_{r}$'s are the same as in the model~\eqref{eq:initialmodel} and the $\epsilon_{rj}$'s follow the assumptions of model~\eqref{eq:initialmodel}. The term $\theta$ represents a random source of variability common to all observations. The $\delta_r$'s, which are the interaction terms, represent random sources of variability unique to each treatment. For instance, in comparing two varieties of wheat, both may respond favorably in going from a drier to moister environment as expressed by the common effect $\theta$, but one variety may respond better than another as expressed through the $\delta_r$'s. The greatest danger to inferring replicability in a changing research environment lies not in those factors that we recognize as having an effect. They often can be accounted for in the model. Rather it is those unknown or unexpected sources of variability that can systematically affect outcomes in unexpected ways. Those are the factors whose effects are captured by the random environmental terms. The  $\delta_r$'s are assumed to be distributed as iid $N(0, \sigma^2_\delta)$, and $\theta$ is assumed to be distributed as $N(0, \sigma^2_{\theta})$ although its distribution does not figure into the discussion except in Section~\ref{sec:realeer}. The random terms in model~\eqref{eq:followupmodel} are assumed to be mutually independent. 
For brevity, we may use M1 to refer to model~\eqref{eq:initialmodel} and M2 to refer to  model~\eqref{eq:followupmodel}.

\subsection{Test Statistic and Distribution} 

Let us consider testing $H_0\!: \mu_1 - \mu_2 = 0$ against $H_a\!: \mu_1 - \mu_2 \neq 0$. Let $n_h$ denote the harmonic mean of the observations
\begin{equation}
\label{eq:defineharmonicmean}
n_h = \frac{2}{1/n_1 + 1/n_2}.
\end{equation}
The test statistic is
\begin{equation}
\label{eq:defineteststatistic}
T = \frac{\overline Y_1 - \overline Y_2}{S_e\sqrt{2/n_h}}
\end{equation} 
where $\bar Y_r$ the sample mean of the responses for treatment $r$ and $S_e$ is the pooled standard deviation of experimental error. 

The treatment effect size (TES), denoted $\Delta$, is defined by
\begin{equation}
\label{eq:definetes}
\Delta = \frac{\mu_1 - \mu_2}{\sigma_e\sqrt{2}}.
\end{equation}
A variant of TES used in the social sciences excludes the $\sqrt{2}$ factor in the denominator, the sample version of which is Cohen's $d$~\citep{cohen1988statistical}. We include the $\sqrt{2}$ because it simplifies some of the mathematical expressions given below. It is also the standard deviation of the difference between two observations, one from each treatment.  

The environmental effect ratio (EER), denoted $\omega$, is
\begin{equation}
\label{eq:defineeer}
\omega = \frac{\sigma_\delta}{\sigma_e}.
\end{equation}
Both $\Delta$ and $\omega$ are dimensionless quantities that can be interpreted without reference to the scale of measurement of the responses or indeed to the area of application. 
The EER also has a connection with Cohen's $d$.  
Under M2, the variance of the difference between sample means is approximately $2\sigma^2_\delta$ when sample sizes are large. 
If we divide this by the error variance $\sigma^2_e$ , this quantity is $\tau^2$, the variance of Cohen’s population effect size~\citep{higgins2002quantifying}. 
That is, $2\omega^2 = \tau^2$ under M2.

Under M2, it can be shown that 
$T/\sqrt{1 + n_h\omega^2}$ has a noncentral t-distribution with degrees of freedom $df =  n_1 + n_2 - 2$ and noncentrality parameter
\begin{equation}
\label{eq:noncentralityparameter}
\frac{\Delta\sqrt{n_h}}{\sqrt{1 + n_h\omega^2}}.
\end{equation} 
We denote the cumulative distribution function (cdf) of this noncentral $t$-distribution as $G_\Delta(t)$. 
The cdf of $T$ under model (2) is given by
\begin{equation}
\label{eq:definecdft}
P(T \leq t~|~\text{M2} ) = G_\Delta\left(\frac{t}{\sqrt{1+n_h\omega^2}} \right).
\end{equation}

The distribution of $T$ under M1 is obtained by setting $\omega = 0$, and under the null hypothesis  $H_0\!: \mu_1 -\mu_2 = 0$, by setting $\Delta = 0$. 
When $\Delta = 0$, $G_\Delta(t)$ is the cdf of the (central) t-distribution with $df = n_1 + n_2 - 2$, which we denote as $G_0(t)$.  
For large samples, the distribution of $T$ can be approximated by a normal distribution with mean $\Delta \sqrt{n_h}$ and variance $1 + n_h\omega^2$. 
\section{Replicability Power}
\label{sec:probofrep}

Suppose the researcher plans the two-sample experiment so that the test achieves a sufficiently large power for a pre-specified treatment effect size $\Delta$ and significance level $\alpha$. For the two-sided test the sample size computation does not depend on the direction of the effect, either positive or negative. However, the direction becomes important because of what can happen in the follow-up experiment. While the probability of rejection in the wrong direction, e.g. lower-tail rejection when $\mu_1 > \mu_2$,  is less than $ \alpha/2$ in the initial experiment, it can be as large as $1/2$ in the follow-up experiment as we discuss below. 
Thus, in considering the power of the follow-up experiment, we must not only consider the magnitude of the effect but also its direction. 
Intuitively, \textit{replicability power} is the probability of detecting an effect of a pre-specified magnitude and direction computed under model~\eqref{eq:followupmodel}. 
This discussion would be unnecessary if we were to do one-sided tests with one-sided power functions, but in doing so, we would fail to consider an important measure of what can go wrong with inferences in a follow-up experiment, namely, the probability of getting significance in the wrong direction in a two-sided test.

We assume that $\mu_1 > \mu_2$ in testing $H_0\!: \mu_1 - \mu_2 = 0$ against $H_a\!: \mu_1 - \mu_2 \neq 0$. Let $t_{\alpha/2}$ denote the $1-\alpha/2$ quantile of the $t$-distribution with $df = n_1 + n_2 - 2$.  With $T \geq t_{\alpha/2}$ being the right direction for rejection in the initial experiment, we obtain from~\eqref{eq:definecdft} that the replicability power, denoted by $p_{rep}$, is 	
\begin{equation}
\label{eq:probofrep}
p_{rep} = P(T \geq t_{\alpha/2}~|~\text{M2}) = 1-G_\Delta\left(t_{\alpha/2}/\sqrt{1+n_h\omega^2}\right),
\end{equation}
and for large $n_1, n_2 $
\begin{equation}
\label{eq:asympprobofrep}
p_{rep}\approx 1 - \Phi\left(\frac{z_{\alpha/2} - \Delta\sqrt{n_h}}{\sqrt{1+n_h\omega^2}} \right)
\end{equation}
where $\Phi$ is the cdf of the standard normal distribution and $z_{\alpha/2}$ is the $1-\alpha/2$ quantile of the standard normal distribution. From expressions~\eqref{eq:definecdft},~\eqref{eq:probofrep}, and~\eqref{eq:asympprobofrep}, we can see that $p_{rep} \rightarrow 1/2$  and $P(T \leq -t_{\alpha/2}) \rightarrow 1/2$ as $\omega \rightarrow \infty$. That is, the replicability power can be as small as 1/2 for large EER regardless of the level of significance and power of the initial experiment, and the probability of an inference in the wrong direction can be as large as 1/2.

\subsection{The Effect of EER on Replicability Power}  
The examples below demonstrate how EER affects replicability power both in small and large samples and for varying effect sizes and significance levels.   

\subsubsection{Example 1: Small \texorpdfstring{$n$}{n} and Traditional \texorpdfstring{$\alpha$}{alpha} \label{subsec:ex1}}

\citet{snedecor1980statistical} illustrate the independent sample $t$-test with data from a study to compare the comb weights of male chicks given one of two hormone treatments. The sample sizes are $n_1 = n_2 =11$, the sample means are $97$ and $56$, and the pooled standard deviation is $12.14$. A test for differences of means gives $p = .003$ for a two-sided test. The observed treatment effect size is 1.02, and a 95\% confidence interval for the true TES is $(.32, 1.72)$~\citep{kadel2012sas}. 

\begin{figure}[ht]
	\centering
	\includegraphics[width = \textwidth]{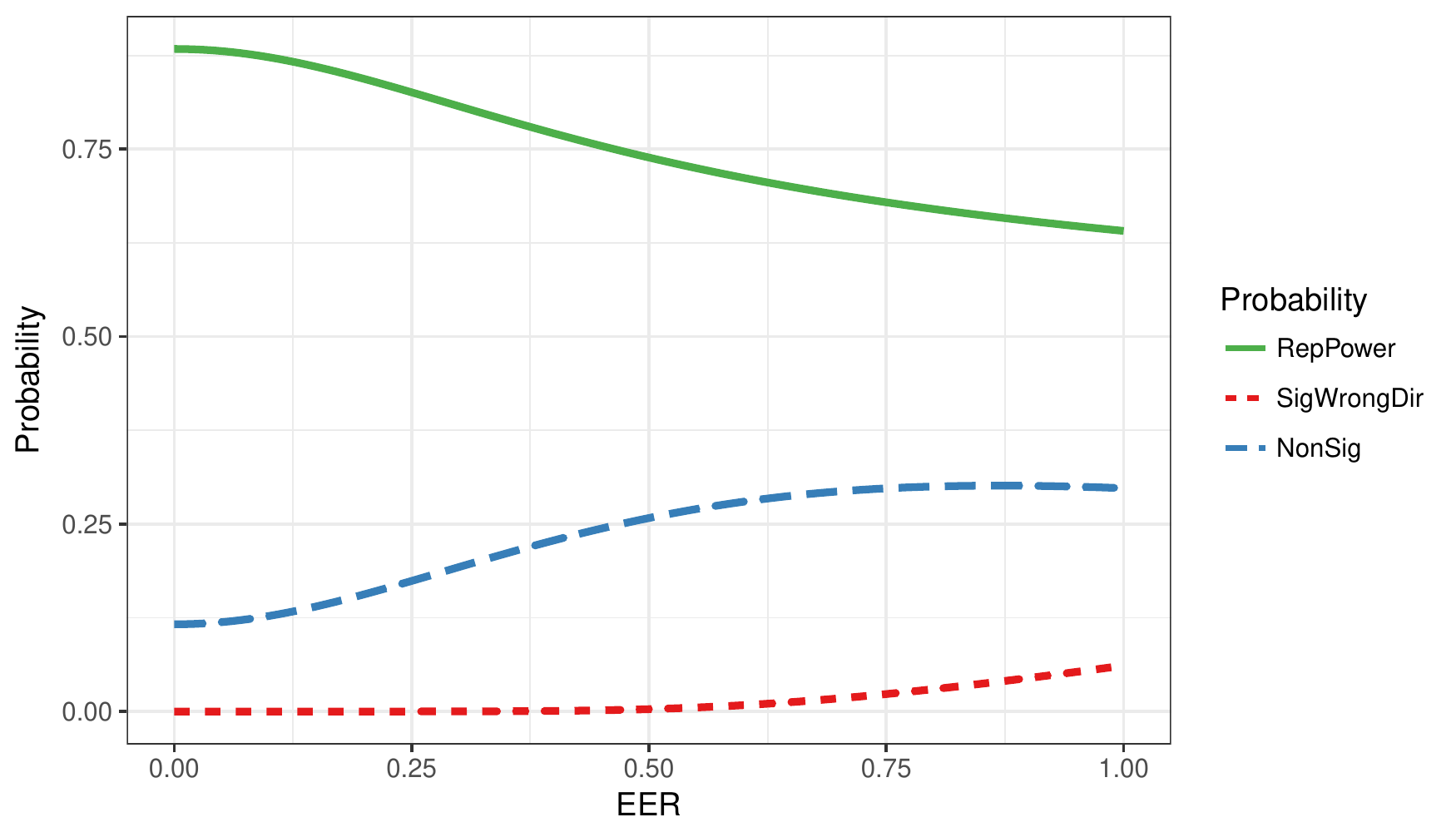}
	\caption{Replicability power, probability of significance in the wrong direction, and probability of non-significance vs.~the environmental effect ratio (EER).  
		Here, the treatment effect size $\Delta = 1.0$, $n_1 = n_2 = 11$, and $\alpha = .05$.  
		The power of the initial study is $.88$, but the power of follow-up studies may be significantly lower when EER is large.}
	\label{figure1}
\end{figure}

To illustrate the sensitivity of the replicability power to EER, suppose that follow-up experiment like this one is planned in which $n_1 = n_2 = 11$, $\alpha = .05$, and true treatment effect size that is of importance is  $\Delta = 1.0$.  
The $.975$ quantile of the $t$-distribution with $df = 20$ is $2.086$, so the replicability power is $p_{rep} = P(T \geq 2.086 ~|~ \text {M2})$.
Figure~\ref{figure1} shows plots of $p_{rep}$, probability of significance in the wrong direction, and probability of non-significance as functions of EER. If the EER $\omega = 0$, then $p_{rep}$ is $.88$ which is the same as the power of the initial test for $\Delta = 1.0$.  
Small values of EER ensure that $p_{rep}$ is large in the follow-up experiment. However, as EER increases, $p_{rep}$ decreases. For instance, if EER= .5, then $p_{rep} = .74$ and the probability of non-significance increases to $.26$. The probability of significance in the wrong direction is negligible in this instance but increases with increasing EER.  


\subsubsection{Example 2: Large \texorpdfstring{$n$}{n} and Small \texorpdfstring{$\alpha$}{alpha}\label{subsec:ex2}}

An initial experiment with a large sample size and small significance level has a low probability of a false positive and a high probability of detecting a treatment effect size of practical importance. 
Thus, large $n$ and small $p$ are often recommended as a prescription for solving replicability problems as discussed in~\citet{singh2017big}. 
However, this can fail in the face of a changing research environment. Table~\ref{tab:largen} gives an example in which $n_1 = n_2 = 300$, $\alpha =.005$, and $\Delta = .25$ or $1.0$. The power of the initial test is $.94$ when $\Delta = .25$, but in a  follow-up experiment with the same TES in which  EER = .5, the replicability power is just $.57$, and the probabilities of finding a non-significant result and significance in the wrong direction are $.22$ and $.21$, respectively. 
On the other hand, if $\Delta = 1.0$, the replicability power is $.95$, and the probabilities of finding non-significance and significance in the wrong direction are negligible. 

\begin{table}
	\begin{center}
		\begin{tabular}{|l| c| c|}\cline{2-3}
			\multicolumn{1}{c|}{}& $\Delta = 0.25$ & $\Delta = 1.00$ \\ \hline
			Power of initial test: Model~\eqref{eq:initialmodel} & 0.93 & 1.00 \\
			Replicability power: Model~\eqref{eq:followupmodel}  & 0.57 & 0.95 \\
			Prob. significant in wrong direction & 0.21 & 0.01 \\
			Prob. non-significant & 0.22 & 0.04\\ \hline			
		\end{tabular}
	\end{center}
	\caption{Replicability power for treatment effect size $\Delta = .25, 1.0$.  Probabilities are computed setting the environmental effect ratio $\omega = 0.5$, $n_1 = n_2 = 300$, and $\alpha = .005$.}
	\label{tab:largen}
\end{table}

\begin{figure}[ht]
	\centering
	\includegraphics[width = \textwidth]{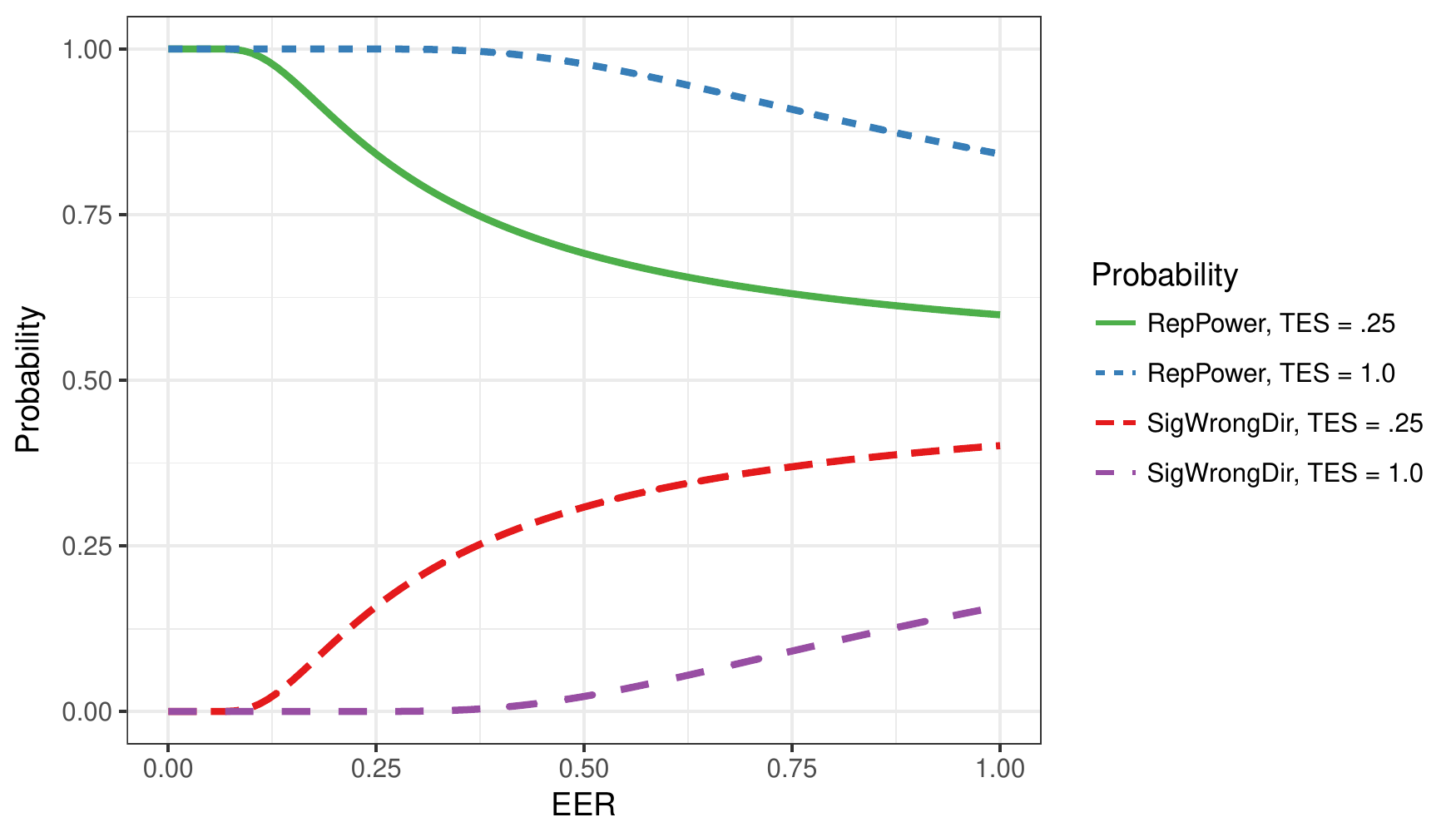}
	\caption{Limiting replicability power and significance in the wrong direction vs.~the environmental effect ratio (EER) $\omega$ for treatment effect size (TES) $\Delta = .25$ and  $1.0.$. 
		Replicability power deviates from 1.0 quickly as the ratio $\Delta/\omega$ decreases. }
	\label{figure2}
\end{figure}

Additional insight can be gained by looking at the limiting case as $n_1, n_2 \rightarrow \infty$.
Under the initial model M1, the power approaches 1 as $n_r \rightarrow \infty$ for any non-zero TES,but this is not the case for the replicability power under model M2.
Because the distribution of the sample means depends on the $\epsilon_{rj}$'s through the standard error $\sigma_e/\sqrt{n_r}$, which is negligible for large samples, the distribution of $T$ in the limit depends only on the distribution of the $\delta_r$'s and is independent of the level of significance. 
Expressed in terms of $\Delta/\omega$ , the replicability power in the limit is given by \begin{equation}
\label{eq:probreplim}
\lim_{n_h\to \infty} P(T \geq t_{\alpha/2} ~|~ \text{M2}) = \Phi(\Delta/\omega),
\end{equation}
and the probability of significance in the wrong direction is $1-\Phi(\Delta/\omega)$.  For instance, if $\Delta = .25$ and $\omega = .5$, the limit is $.69$, which is the largest that the replicability power can be as a function of sample size for these values of $\Delta$ and $\omega$, and the probability of significance in the wrong direction is $.31$. However, if $\Delta = 1$, these limiting values are .98 for power and .02 for probability of significance in the wrong direction. Thus, as expected, larger values of TES help mitigate the negative effects of EER on replicability power. This is further  is illustrated in Figure~\ref{figure2} which has plots of the limiting replicability power in~\eqref{eq:probreplim} and probability of significance in the wrong direction vs EER for $\Delta = .25$ and $1.0$. 


\subsection{Relative Efficiency of the Initial and Follow-up Experiments}

If two tests that are performed at the same level of significance have the same power for a given effect size, then the ratio of their respective sample sizes is a measure of the relative efficiency of the two tests. 
We adapt this idea to obtain the relative efficiency of the $t$-test when applied to the initial and follow-up experiments. 
Suppose $\mu_1 > \mu_2$ as before and suppose, for a given $\Delta$ and $\alpha$, we would like the replicability power to be the same as the power of the initial test. Let $n_I$ and $n_F$ be sample sizes per treatment in the initial and follow-up experiments, respectively, so that the initial and replicability powers are the same. The relative efficiency of the follow-up experiment to the initial experiment is $n_I/n_F$. For simplicity, we use the normal approximation in~\eqref{eq:asympprobofrep}. 

The sample size $n_I$ necessary for the initial experiment to have power $1-\beta$  for treatment effect size $\Delta$  is $n_I = (z_{\alpha/2} + z_{\beta})^2/\Delta^2$. 
For the follow-up experiment, $n_F$ can be found by setting the replicability power~\eqref{eq:asympprobofrep} to $1 - \beta$ and solving for $n_F$. 
This is equivalent to solving for $n_F$ in the formula

\begin{equation}
\label{eq:findnmodel2}
z_{\alpha/2} + z_\beta\sqrt{1+n_F\omega^2} = \Delta \sqrt{n_F}.
\end{equation}
It is possible that there is no value of $n_F$ that will satisfy~\eqref{eq:findnmodel2}, in which case, the relative efficiency is $0$. 
\begin{figure}[ht]
	\centering
	\includegraphics[width = \textwidth]{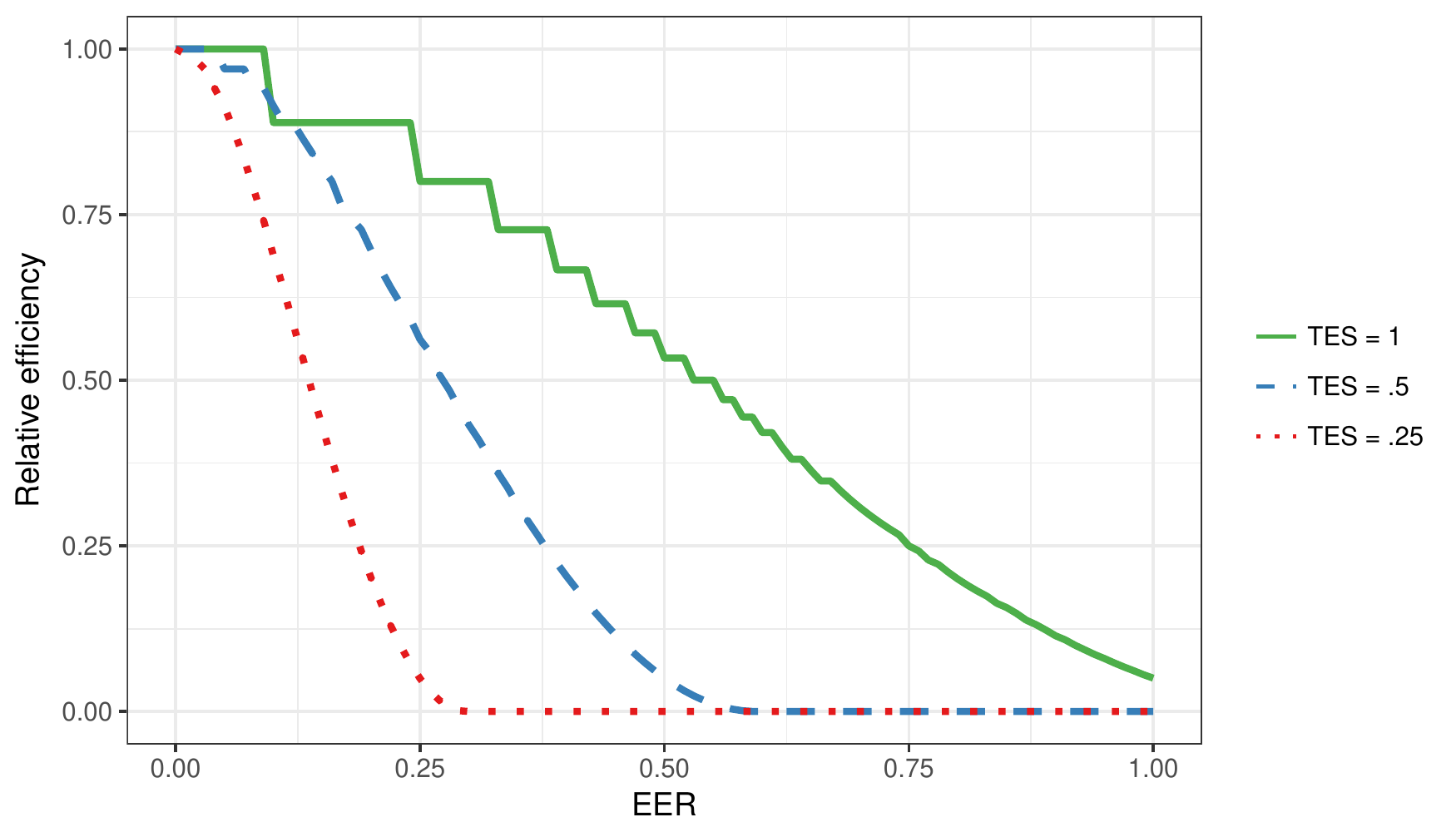}
	\caption{Relative efficiency in sample sizes between the initial and follow up experiments vs.~environmental effect ratio (EER) $\omega$ for varying treatment effect sizes (TES) $\Delta$.  Here, $\alpha = .05$ and $1-\beta = .8$.  Relative efficiency decreases rapidly as the ratio $\Delta/\omega$ decreases.}
	\label{figure3}
\end{figure}

Figure~\ref{figure3} shows plots of the relative efficiencies against EER for values of the TES $\Delta = .25$ and $.5$ given level of significance $\alpha = .05$ and power of the initial experiment set at $.8$.  
If $\Delta = .25$ and $\omega = .2$, the sample sizes are $n_I = 126$ and $n_F = 626$ with relative efficiency of only $.20$ (computed sample sizes are rounded up to the nearest integer). 
Relative efficiency increases with increasing $\Delta$. 
For instance, if $\Delta = .5$ when $\omega$ = .2, the sample sizes are $n_I = 32$ and $n_F = 46$, giving a relative efficiency of .70. 

A researcher who wishes to have the results of the initial experiment confirmed by a follow-up experiment is at a disadvantage if the follow-up experiment is done at the same sample size as the initial experiment. 
The loss of efficiency in the follow-up experiment can be substantial. 
One may be able to compensate by adjusting the sample size of the follow-up experiment to put it on an equal footing with the initial experiment, although this is not always possible. 
As with all sample size determinations, prior knowledge of the size of the variance components is required to determine the sample size for the follow-up experiment.

\section{Assessing Replicability through Varying EER}
\label{sec:adjpval}

We now discuss the sensitivity of inferences to changes in environment through evaluating inferential quantities across a range of hypothetical values for the EER. We first derive, what we call, \textit{broad-inference} $p$-values and confidence levels that take into account the environmental variability present in the follow-up experiment.
These inferential quantities are functions of the data and the unknown EER parameter $\omega$.
We then demonstrate how evaluating these quantities across a spectrum of hypothetical EER values can aid researchers in assessing the replicability of an experiment, and in particular, help determine how much environmental variability may be tolerated before results can no longer expect to be replicated.
This analysis can be performed graphically through EER \textit{profile plots}.
The process is similar to the sensitivity analysis approach popularized by~\citet{rosenbaum2002overt}, except results are evaluated under differing magnitudes of unobserved across-experiment variability rather than unobserved confounding.

\subsection{Broad-Inference \texorpdfstring{$p$}{p}-Values and Confidence Levels}

We now derive the broad-inference $p$-values and confidence intervals.

Our derivation is similar to that in~\citet{kafkafi2017addressing} except we write expressions in terms of the unobserved EER $\omega$.
Suppose that a follow-up experiment is performed in a randomly selected environment. 
Consider the conditional distribution of $T$ in that environment given $\delta_1 = \delta^*_1$ and $\delta_2 = \delta^*_2$. 
This distribution depends on $\mu_1 - \mu_2 + \delta^*_1 - \delta^*_2$  instead of only $\mu_1 - \mu_2$ as in M1. 
Thus, an observed value of $T$ that may be judged “extreme” under M1 and $H_0\!: \mu_1 - \mu_2 = 0$ may be either more extreme or less extreme under the follow-up model M2 depending on the size of the (unobserved) $\delta^*_1 - \delta^*_2$. 
Also, the conditional confidence interval in the follow-up experiment is centered on $\mu_1 - \mu_2 + \delta^*_1 - \delta^*_2$  not  $\mu_1 - \mu_2$, so the stated level of confidence for capturing $\mu_1 - \mu_2$ in the interval is smaller under M2 than under M1. 
The broad-inference values are conceptually computed by averaging the conditional $p$-values and confidence levels across the environments, i.e. with respect to the distribution of $\delta_1 - \delta_2$.  
However, we performed the actual computations directly from~\eqref{eq:definecdft}.

\subsubsection{Broad-Inference \texorpdfstring{$p$}{p}-Value}

The observed effect size computed from the initial experiment is defined to be
\begin{equation}
\label{eq:defineobseffsize}
\Delta^* = \frac{\bar y_1 - \bar y_2}{\sqrt 2 s_e}, 
\end{equation}
where the lower-case letters denote observed values from the two treatments. 
We assume $\Delta^* > 0$ consistent with our assumption that $\mu_1 > \mu_2$.  


Because the distribution of $T$ under M2 is given by~\eqref{eq:probofrep} with $\mu_1 - \mu_2 = 0$ and $df = n_1 + n_2 -2$, 
the broad-inference two-sided $p$-value for the effect size $\Delta^*$ in a follow-up experiment is
\begin{equation}
\label{eq:adjpvalue}
P\left(|T| > \Delta^*\sqrt{n_h} ~|~ \text{M2}, \mu_1 - \mu_2 = 0\right)
= 2\left(1- G_0\left(\frac{\Delta^*\sqrt{n_h}}{\sqrt{1+ n_h\omega^2}}\right)\right).
\end{equation} 
The limit of~\eqref{eq:adjpvalue} as $n_h \to \infty$ is the \textit{asymptotic} broad-inference $p$-value, which is
\begin{equation}
\label{eq:asymtpvalue}
2(1-\Phi(\Delta^*/\omega)). 
\end{equation}
Note that~\eqref{eq:adjpvalue} decreases as $n_h$ increases so that the asymptotic broad-inference $p$-value is also the minimum. 
For the asymptotic broad-inference $p$-value to be $\alpha$, we must have $\Delta^* =  z_{\alpha/2}\omega$. 
If $\Delta^*$ is less than this, the broad-inference $p$-value cannot attain a level of significance $\alpha$ regardless of sample size. 

\subsubsection{Broad-Inference Level of Confidence and Confidence Intervals}
Confidence intervals may be preferred to hypothesis tests because they provide more information, but whether one uses confidence intervals or hypothesis tests, the problems posed by the random factors in the follow-up model do not go away. 
The confidence interval $(\bar y_1 - \bar y_2) \pm t_{\alpha/2}s_e\sqrt{2/n_h}$, while having level of confidence $1-\alpha$ for $\mu_1 - \mu_2$ in the initial experiment, has a smaller broad-inference level of confidence.  The derivation of the confidence level begins with the distribution of the statistic

\begin{equation}
\label{eq:defineteststatisticc}
T_\mu = \frac{\overline Y_1 - \overline Y_2-(\mu_1 - \mu_2)}{S_e\sqrt{2/n_h}}
\end{equation}
under M2. 
It can be seen that $T_\mu /\sqrt{1+n_h\omega^2}$ has a t-distribution with $df = n_1 + n_2 - 2$. 
Thus, the broad-inference confidence level is the probability that the confidence interval contains $\mu_1 - \mu_2$  which can be shown to be 
\begin{equation}
\label{eq:adjconflevel}
2 G_0\left( \frac{t_{\alpha/2}}{\sqrt{1+n_h\omega^2}} \right) - 1.
\end{equation}
If   $\omega > 0$, the broad-inference level of confidence approaches 0 as $n_h\to \infty$. 
Thus, for large $n_h$, the traditional confidence interval in a follow-up experiment almost certainly will not contain $\mu_1 - \mu_2$.

Additionally, we can manipulate the above equation to find an broad-inference confidence interval for a given significance level $\alpha$. A \textit{broad-inference} $1-\alpha$ confidence interval is
\begin{equation}
\label{eq:adjconfint}
(\bar y_1 - \bar y_2) \pm t_{\alpha/2}s_e\sqrt{2/n_h + 2\omega^2}.
\end{equation}
Of note, as $n_h\rightarrow \infty$, the length of the confidence interval for a fixed confidence level $1-\alpha$ approaches $\omega\sqrt{2}$.
That is, when accounting for the presence of the EER, the confidence interval in~\eqref{eq:adjconfint} no longer is expected to converge to the true value of the difference in means.

\subsection{Assessing Replicability through EER Profile Plots for Example 1}
We now use EER profile plots to assess the replicability of the experiment performed by~\citet{snedecor1980statistical} on the use of in Example 1 in Section~\ref{subsec:ex1}.
These are obtained by taking the quantities in~\eqref{eq:adjpvalue} and~\eqref{eq:adjconflevel} and plotting these values against varying hypothetical values of the EER $\omega$.
These plots allow for analysis on how sensitive results are to environmental variability.

\begin{figure}[ht]
	\centering
	\includegraphics[width = \textwidth]{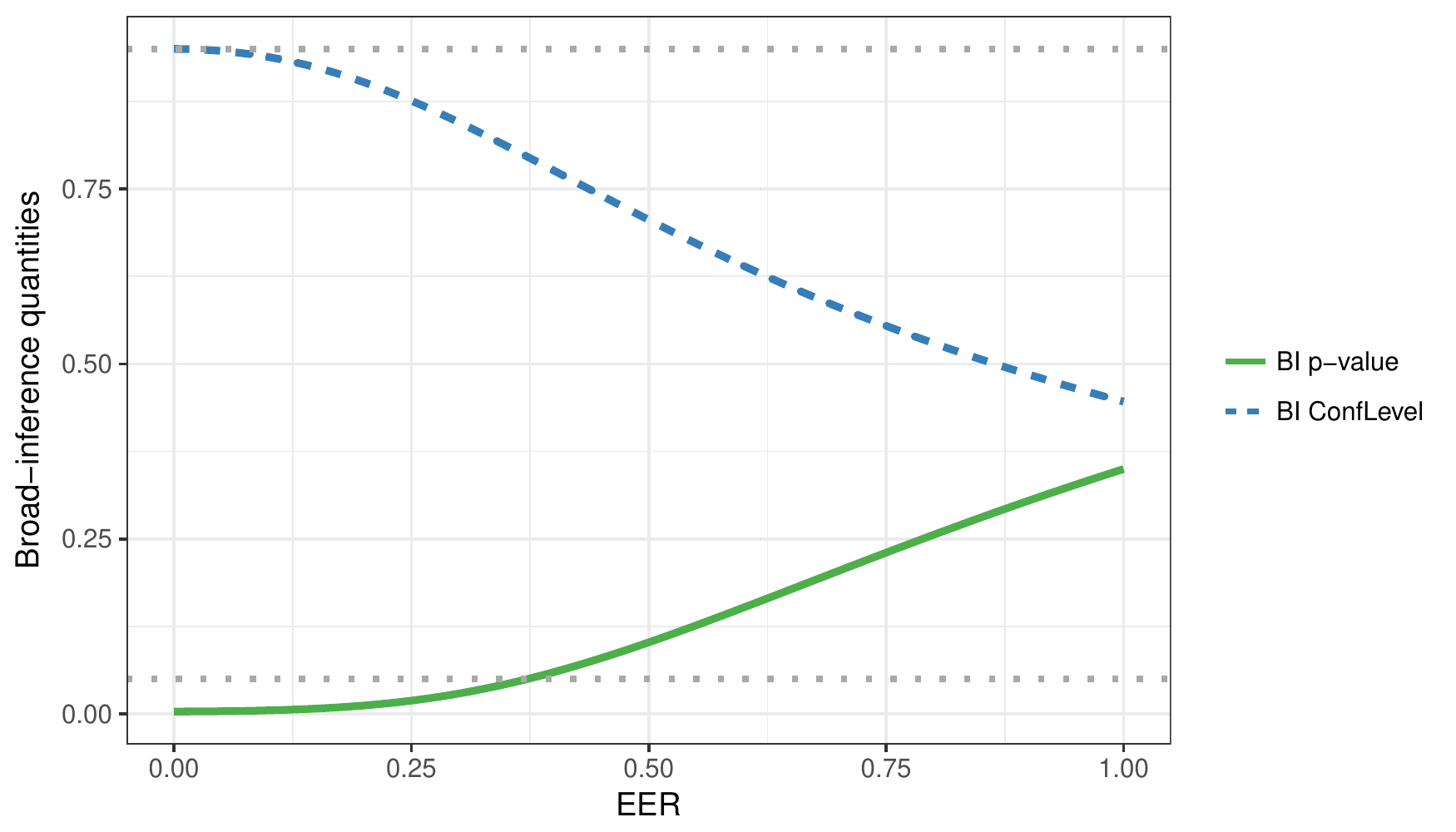}
	\caption{Profile plot of broad-inference (BI) $p$-values and confidence levels vs.~environmental effect ratio (EER) for observed treatment effect size $\Delta^{*}=1.0$ and $n_1 = n_2 = 11$.  
		The dotted gray line at the bottom denotes $p=.05$ and at the top denotes the initial confidence level $1-\alpha = .95$.  
		The initial $p$-value of the study (when EER $\omega = 0$) is $p = .003$.}
	\label{figure4}
\end{figure}

Figure~\ref{figure4} gives these profile plots.
Conventional analysis yields a difference in hormone treatments ($p = .003$) that would typically be indicative of strong statistical significance. 
However, this $p$-value only holds under the environment in which the experiment was conducted.


Indeed, if $\omega$ is quite small, inferences made using broad-inference $p$-values will be the same as those from traditional statistical tests.
However, if $\omega > .38$, which seems to be a reasonable value of the  environmental variability (see Section~\ref{sec:realeer}), the broad-inference $p$-value would no longer indicate significance at the $5\%$ level and the broad-inference level of confidence would fall below $.8$ from an initial value of $.95$. 
If $\omega > .65$, which is plausible for experiments that have environments that are difficult to regulate, the broad-inference $p$-value is  $p = .178$, for which only the most unscrupulous of practitioners would claim any kind of significance (for examples, see~\citet{hankins2013still}).
That is, despite seemingly strong statistical significance of treatment under conventional analysis, there seems to be weak evidence that the result is actually due to treatment efficacy rather than the variability that may exist across different experimental environments.
A researcher interested in performing a follow-up study to this experiment should not be shocked if the original result failed to replicate. 

\section{Plausible values of EER}
\label{sec:realeer}

Estimates of EER from multi-environment experiments in four different areas of application were obtained to show plausible values of EER in practice.  
With such information from prior data or experiences, a researcher can place plausible bounds on EER from which bounds can be placed on replicability power, sample sizes in replicability studies, broad-inference $p$-values, and confidence levels, and help interpret and make conclusions from EER profile plots. 

\subsection{Example 3: Genetics}
\label{subsec:ex3esteer}
\citet{Kafkafi:2005} used a mixed model to decompose the total variance of 17 endpoints in a multi-laboratory experiment into the following components: between-genotype, between-laboratory, genotype by laboratory interaction, and within-laboratory. 
We calculated EER for these endpoints from the proportions of total variance for interaction and error presented in graphical form in \citet{Kafkafi:2005} and in table form in a supplement to their paper. Results are shown in Table~\ref{Table.2}. 
Values of $\omega$ range from 0 to .63 with a median of .34. 

\begin{table}[ht]
	\centering
	\begin{tabular}{|c| c| l| c| } 
		\hline
		\multicolumn{4}{|c|}{EER for different endpoints \citep{Kafkafi:2005}} \\
		\hline
		\multicolumn{1}{|l|}{Endpoint(Response)} &{EER}&{Endpoint(Response)} & {EER}\\
		\hline
		\multicolumn{1}{|l|}{lingering time} &{0.63}&{distance traveled} &{0.57}\\
		\hline
		\multicolumn{1}{|l|}{segment max speed} &{0.40}&{excursions} &{0.37}\\
		\hline
		\multicolumn{1}{|l|}{time for tum} &{0.13}&{radius of tum} &{0.34}\\
		\hline
		\multicolumn{1}{|l|}{segment length} &{0.51}&{center time} &{0.40}\\
		\hline
		\multicolumn{1}{|l|}{progression segments} &{0.14}&{segment acceleration} &{0.38}\\
		\hline
		\multicolumn{1}{|l|}{homebase occupancy} &{0.0}&{lingering mean speed} &{0.35}\\
		\hline
		\multicolumn{1}{|l|}{diversity} &{0.0}&{stops per excursion} &{0.18}\\
		\hline
		\multicolumn{1}{|l|}{lingering spatial spread} &{0.19}&{relative activity decrease} &{0.03}\\
		\hline
		\multicolumn{1}{|l|}{latency to half max speed} &{0.0}&{} &{}\\
		\hline
	\end{tabular}
	\caption{Estimated environmental effect ratios (EER) for a multi-lab experiment.}
	\label{Table.2}
\end{table}

\subsection{Example 4: RCB in Agriculture}

In agriculture experimentation, there is often non-homogeneity in the environment that is accounted for by blocking in the design of the experiment. If the treatments appear more than once in each block, then the usual RCB analysis with fixed effect ``treatment'' and random effects ``block'' and ``block$\times$treatment'' would give us an estimate of the component of variance $\sigma^2_\delta$ for the random interaction of blocks and treatments. If we regard the block itself as an ``environment,'' then the estimate of the ratio $\sigma_\delta/\sigma_e$ will suggest a plausible value for EER. 
We analyzed data from~\citet[p.~267]{snedecor1980statistical} using a randomized block design in which there are 3 treatments measured each of 4 times in each of 5 blocks.
The data are the number of wireworms in soil samples treated with either one of two fumigants or a control. We obtained the estimates of the components of variance as shown in Table~\ref{tab:ex3ests} and computed the estimate of $\omega$ to be $.65$.  

Note that the ``block$\times$treatment''mean square is the error mean square for testing treatment effects, not the residual. 
If data were taken from only one block, the residual would be the wrong error term for making inferences across blocks. 

\begin{table}
	\begin{center}
		\begin{tabular}{|l| c| c| c | c | }\hline
			Source of Var. & Estimate & SE & $Z$-Value & $Pr > Z$ \\ \hline
			blk & $1.1052$ & $2.4502$ & $0.45$ & $0.326$ \\ 
			blk$\times$trt $(=\sigma^2_\delta)$ & $3.8559$ & $3.1035$ & $1.24$ & $0.107$ \\
			Residual $(= \sigma^2_e)$ & $9.1056$ & $1.9196$ & $4.74$ & $<0.0001$\\
			$\text{EER} = \sigma_\delta/\sigma_e$ & 0.6507 & & &\\
			\hline 	
		\end{tabular}
	\end{center}
	\caption{Variance component estimates for Example 4.}
	\label{tab:ex3ests}
\end{table}

\subsection{Example 5: EER via Intraclass Correlation, Horticulture and Education }

We may relate EER to the \textit{intraclass correlation} from which a bound may be placed on EER. If we randomly select an environment then take observations according to M2, the observations within treatment are correlated because of the random term $\theta + \delta_i$ which is common to all the observations within the treatment. 
The intraclass correlation is given by

\begin{equation}
\label{eq:deficc}
\rho = \frac{\sigma^2_\theta + \sigma^2_\delta}{\sigma^2_\theta + \sigma^2_\delta + \sigma^2_e} .
\end{equation}
This quantity is smallest when $\sigma^2_\theta = 0$, so $\rho > \sigma^2_\delta/(\sigma^2_\delta + \sigma^2_e)$. 
It follows that
\begin{equation}
\label{eq:boundomega}
\omega < \sqrt{\rho/(1-\rho)} .
\end{equation}We use this method of bounding EER in data taken from horticulture and education. 

The value of $\rho$ was estimated in~\citet{perrett2006method} for eight cultivars inoculated with spider mites in 4 greenhouses which are the environments for this example. The values of estimated $\rho$ ranged from $0$ to $.30$ with a median of $.12$. If $\rho = .30$, then $\omega < .65$, and if $\rho = .12$, then $\omega < .37$. 

We also compute bounds on EER from data obtained by~\citet{perrett2004using}. 
Here, the intraclass correlation was for 14 multi-section university courses using data from the fall and spring semesters of the years 2001-2003 for a total of 43 course-semester combinations. 
The data were grades based on a 4 point scale. The largest value of $\rho$ was .34 ($\omega < .72$). 
However, the remaining 42 values of $\rho$ were .23 or less ($\omega < .55$) and the median was .07 ($\omega < .27$).

\section{Summary and Conclusions \label{sec:conclude}}

It would be foolhardy to perform an experiment at a research facility if it were believed that the results would apply only to that facility at the time the research is done and under the conditions that prevail at that time. 
Yet the methods that we teach, the software that we use, and the consulting advice that we provide often allow for formal statistical inferences to be made only under those circumstances. 
While the replicability crisis has led to extensive commentary on the improper application of statistical methods, in particular, the uncritical use of ``$p < .05$,'' problems of replicability cannot be fully addressed without considering the effect that the research environment has on outcomes and inferences. This is true whether the inferences are frequentist or Bayesian and whether we use $p < .05$ or abandon significance tests altogether.

By using a mixed model and EER as a measure of the size of the environmental effect for a follow-up experiment, one can quantitatively assess the effect that the research environment has on power, sample size selection, $p$-values, and confidence levels. The usual $p$-values and confidence levels can be misleading in answering questions of replicability if EER $\omega > 0$. Broad-inference $p$-values and confidence levels are more conservative and more appropriate for this purpose because they account for potential changes in the research environment when an experiment is redone. 

A large treatment effect size, which like EER is dimensionless, can help mitigate the negative effects of environmental factors, but small treatment effect sizes, even if statistically significant, should be viewed with caution. 
For instance, a treatment effect size less than $.7$ (Cohen’s d less than .5) could not attain a broad-inference $p$-value less than .05 regardless of sample size when EER is greater than $.36$, which is a plausible value for EER in practice. 
In areas where the effects of the research environment are difficult to control or the effect sizes tend to be small, the only recourse the researcher will have to separate treatment effects from random error would be to do the experiment in several, perhaps many, randomly selected environments. 
EER profile plots of broad-inference $p$-values and broad-inference confidence levels across hypothetical values of EER and plots of replicability power against EER can be used to judge the sensitivity of results to changing environments and to assist in making judgments about the replicability of experimental outcomes.  
This is especially helpful when it is impractical to redo an experiment or when one wants to assess likely replicability before an experiment can be redone.

In planning a follow-up experiment, the replicability power will be less than that of the initial experiment for the same sample size if the initial power is greater than $1/2$ and $\omega > 0$.  
Thus, the sample size must be greater in the follow-up experiment for the two experiments to be on an equal footing. Large $n$ alone cannot control replicability power which can be as small as $1/2$ when EER is large relative to treatment effect size.  
Similarly, the probability of obtaining significance in the wrong direction can be as large as $1/2$ again depending on the size of EER.

There are implications for teaching as well. 
For instance, in teaching the two-sample $t$-test in introductory statistical methods courses, we dutifully note that the observations are randomly selected from normally distributed populations and caution students that inferences only apply to these populations.
What may be ignored or given only slight attention is the fact that inferences are also limited by the conditions that prevail at the time the data are taken.
If the objective of statistics is ``better scientific investigation'' as~\citet{box1990commentary} asserted, then it is incumbent upon the teacher to give prominence to this point. 
The limitations on inferences imposed by the research environment should be as much of a concern as the distributional assumptions of the methods used to analyze data. 

Greater effort needs to be made to get estimates of components of variance due to environmental effects and in particular EER which plays a key role in replicability. 
Where available, these should be reported along with means, standard errors, and effect sizes in research results. 
With greater knowledge of EER in various contexts, it will become possible to make more informed judgments about the potential replicability of results from a single experiment.

\bibliographystyle{agsm}

\bibliography{reproduce}
\end{document}